\documentclass[11pt,a4paper,showpacs,showkeys,,superscriptaddress]{article}

\usepackage{epsfig}
\usepackage{amsmath,amssymb}
\usepackage{graphicx}
\usepackage{xcolor}
\usepackage{subfigure}

\usepackage{latexsym}
\usepackage{hyperref}
\hypersetup{colorlinks,%
  citecolor=blue,%
  linkcolor=red,%
  pdftex}

\begin{document}

\baselineskip 6mm
\renewcommand{\thefootnote}{\fnsymbol{footnote}}

\newcommand{\nc}{\newcommand}
\newcommand{\rnc}{\renewcommand}



\newcommand{\tcb}{\textcolor{blue}}
\newcommand{\tcr}{\textcolor{red}}
\newcommand{\tcg}{\textcolor{green}}


\def\beq{\begin{equation}}
\def\eeq{\end{equation}}
\def\ba{\begin{array}}
\def\ea{\end{array}}
\def\bea{\begin{eqnarray}}
\def\eea{\end{eqnarray}}
\def\nn{\nonumber}


\def\CMP{Commun. Math. Phys.~}
\def\JHEP{JHEP~}
\def\Pre{Preprint}
\def\PRL{Phys. Rev. Lett.~}
\def\PR {Phys. Rev.~}
\def\CQG {Class. Quant. Grav.~}
\def\PL {Phys. Lett.~}
\def\NP {Nucl. Phys.~}

\def\G{\Gamma}

\def\S{{\bf S}}
\def\C{{\bf C}}
\def\Z{{\bf Z}}
\def\R{{\bf R}}
\def\N{{\bf N}}
\def\M{{\bf M}}
\def\P{{\bf P}}
\def\bm{{\bf m}}
\def\bn{{\bf n}}

\def\CA{{\cal A}}
\def\CB{{\cal B}}
\def\CC{{\cal C}}
\def\CD{{\cal D}}
\def\CE{{\cal E}}
\def\CF{{\cal F}}
\def\CH{{\cal H}}
\def\CM{{\cal M}}
\def\CG{{\cal G}}
\def\CI{{\cal I}}
\def\CJ{{\cal J}}
\def\CL{{\cal L}}
\def\CK{{\cal K}}
\def\CN{{\cal N}}
\def\CO{{\cal O}}
\def\CP{{\cal P}}
\def\CQ{{\cal Q}}
\def\CR{{\cal R}}
\def\CS{{\cal S}}
\def\CT{{\cal T}}
\def\CU{{\cal U}}
\def\CV{{\cal V}}
\def\CW{{\cal W}}
\def\CX{{\cal X}}
\def\CY{{\cal Y}}
\def\CZ{{\cal Z}}

\def\We{{W_{\mbox{eff}}}}


\newcommand{\Lie}{\pounds}

\newcommand{\p}{\partial}
\newcommand{\bp}{\bar{\partial}}

\newcommand{\half}{\frac{1}{2}}

\newcommand{\bfalpha}{{\mbox{\boldmath $\alpha$}}}
\newcommand{\bfbeta}{{\mbox{\boldmath $\beta$}}}
\newcommand{\bfgamma}{{\mbox{\boldmath $\gamma$}}}
\newcommand{\bfmu}{{\mbox{\boldmath $\mu$}}}
\newcommand{\bfpi}{{\mbox{\boldmath $\pi$}}}
\newcommand{\bfvarpi}{{\mbox{\boldmath $\varpi$}}}
\newcommand{\bftau}{{\mbox{\boldmath $\tau$}}}
\newcommand{\bfeta}{{\mbox{\boldmath $\eta$}}}
\newcommand{\bfxi}{{\mbox{\boldmath $\xi$}}}
\newcommand{\bfkappa}{{\mbox{\boldmath $\kappa$}}}
\newcommand{\bfepsilon}{{\mbox{\boldmath $\epsilon$}}}
\newcommand{\bfTheta}{{\mbox{\boldmath $\Theta$}}}

\newcommand{\bz}{{\bar{z}}}

\newcommand{\dalpha}{\dot{\alpha}}
\newcommand{\dbeta}{\dot{\beta}}
\newcommand{\blambda}{\bar{\lambda}}
\newcommand{\btheta}{{\bar{\theta}}}
\newcommand{\bsigma}{{{\bar{\sigma}}}}
\newcommand{\bepsilon}{{\bar{\epsilon}}}
\newcommand{\bpsi}{{\bar{\psi}}}


\def\ct{\cite}
\def\la{\label}
\def\eq#1{(\ref{#1})}


\def\a{\alpha}
\def\b{\beta}
\def\g{\gamma}
\def\G{\Gamma}
\def\d{\delta}
\def\D{\Delta}
\def\ep{\epsilon}
\def\e{\eta}
\def\ph{\phi}
\def\Ph{\Phi}
\def\ps{\psi}
\def\Ps{\Psi}
\def\k{\kappa}
\def\l{\lambda}
\def\L{\Lambda}
\def\m{\mu}
\def\n{\nu}
\def\th{\theta}
\def\Th{\Theta}
\def\r{\rho}
\def\s{\sigma}
\def\S{\Sigma}
\def\ta{\tau}
\def\o{\omega}
\def\O{\Omega}
\def\pr{\prime}
\def\f{\varphi}


\def\half{\frac{1}{2}}

\def\goto{\rightarrow}

\def\na{\nabla}
\def\grad{\nabla}
\def\curl{\nabla\times}
\def\div{\nabla\cdot}
\def\pa{\partial}

\def\bra{\left\langle}
\def\ket{\right\rangle}
\def\lb{\left[}
\def\lc{\left\{}
\def\ls{\left(}
\def\lp{\left.}
\def\rp{\right.}
\def\rb{\right]}
\def\rc{\right\}}
\def\rs{\right)}
\def\cl{\mathcal{l}}

\def\vac#1{\mid #1 \rangle}

\def\td#1{\tilde{#1}}
\def\check{ \maltese {\bf Check!}}


\def\Tr{{\rm Tr}\,}
\def\det{{\rm det}\,}


\def\bc#1{\nnindent {\bf $\bullet$ #1} \\ }
\def\ch {$<Check!>$ }
\def\ss {\vspace{1.5cm}}

\begin{titlepage}

\hfill\parbox{5cm} { }

\hskip1cm

\vspace{10mm}

\begin{center}
{\Large \bf Scaling symmetry and scalar hairy Lifshitz black holes}

\vskip 1. cm
  { Seungjoon Hyun\footnote{e-mail : sjhyun@yonsei.ac.kr}, Jaehoon Jeong\footnote{e-mail : jjeong@physics.auth.gr}, Sang-A Park\footnote{e-mail : sangapark@yonsei.ac.kr},
  Sang-Heon Yi\footnote{e-mail : shyi@yonsei.ac.kr} 
  }

\vskip 0.5cm

{\it Department of Physics, College of Science, Yonsei University, Seoul 120-749, Korea}\\
{\it ${}^{\dagger}$Institute of Theoretical Physics,
Aristotle University of Thessaloniki,
54124, Thessaloniki, Greece}
\end{center}

\thispagestyle{empty}

\vskip1.5cm

 
\centerline{\bf ABSTRACT} \vskip 4mm
 \vspace{1cm} 
\noindent  By utilizing the scaling symmetry of the reduced action for planar black holes, we  obtain the corresponding conserved charge. We use the conserved charge to find  the generalized Smarr relation of  static hairy planar black holes in various dimensions. Our results not only reproduce the relation in the various known cases but also give the new relation in the  Lifshitz planar black holes with the scalar hair. 
\vspace{2cm}


\end{titlepage}

\renewcommand{\thefootnote}{\arabic{footnote}}
\setcounter{footnote}{0}

\section{Introduction}

Black holes have been fascinating  objects which still await their full understanding.  In the asymptotically flat case, one of  interesting aspects of black holes   is the statement that there cannot be extra matter profiles outside the black hole horizon  except the one responsible for electromagnetic charges, which is  dubbed as no-hair theorem. This property of black holes is widely recognized as the cornerstone of the information problem in black hole physics.  Since the explicit formulation of no-hair theorem is usually made only in four dimensional asymptotically flat space~\cite{Israel:1967wq,Carter:1971zc,Bekenstein:1971hc,Robinson:1975bv} (however, see~\cite{Herdeiro:2014goa,Herdeiro:2015gia,Herdeiro:2015waa} for the possibility of  black holes with non-stationary scalar hairs), it is an interesting question to understand what happens in the space of other asymptotic structure or in the space of other than four dimensions. 

In the case of the asymptotically AdS spacetime  a scalar field can have a negative mass square  while maintaining the unitarity of the scalar field. This indicates that a scalar field in the asymptotically AdS spacetime may behave differently from the one in the asymptotically flat spacetime. Indeed, there are numerous analytic examples which have a scalar hair in the asymptotically AdS spacetime~\cite{Henneaux:2002wm,Henneaux:2004zi,Henneaux:2006hk,Hertog:2004dr,Anabalon:2013qua,Faedo:2015jqa}. Furthermore, the complex scalar field outside the black hole horizon is one of the essential ingredients for the construction  of the holographic superconductor model~\cite{Hartnoll:2008kx,Hartnoll:2008vx}. This gives us an impression that  the theorem may be regarded as a specific property of the asymptotically flat spacetime and is nullified in the asymptotically AdS spacetime. Nevertheless, there have been some attempts to extend no hair theorem to the asymptotically AdS spacetime by imposing some conditions on the scalar field potential~\cite{Hertog:2006rr}. The existence of various  analytic black hole solutions with a scalar hair shows us that those solutions evade some conditions of no-hair theorem.

At the present stage of the investigation on scalar hairy black holes, it seems very useful to explore the properties of scalar hairy black holes in a generic setup resorting to neither specific solutions nor the form of the scalar potential.  In this regard, there was an interesting observation on the scaling symmetry in the reduced action formalism~\cite{Banados:2005hm}.  Concretely speaking, in three-dimensional Einstein gravity with a minimally-coupled scalar field, it has been observed that the reduced action of spherically symmetric black holes has the novel scaling symmetry and  shown that its conserved charge leads to a Smarr relation even for  scalar hairy black holes. This symmetry and the derivation of the Smarr relation do not use any specific property of the scalar potential. Therefore, it would be interesting to extend this approach to more generic cases.   We will show that the scaling symmetry of the reduced action is not restricted to the asymptotically AdS spacetime nor to Einstein gravity.

In recent years, another asymptotic geometry called as the  Lifshitz space arouses some interests in the  view point of the AdS/CMT correspondence.   Contrary to the AdS space, the Lifshitz space has the  anisotropic scaling of time and space as 
\[   
 t \longrightarrow \lambda^{z} t \,, \qquad x \longrightarrow  \lambda x \,,
\]
where $z$ is called  the dynamical exponent. We would like to study the scalar hairy Lifshitz black holes in the reduced action formalism irrespective of the existence  of the analytic solutions. 
 This space has been studied as the gravity dual to the non-relativistic Lifshitz scaling physical system~\cite{Kachru:2008yh,Son:2008ye}.  Analytic Lifshitz black hole solutions in three dimensions have been found  and the Smarr relation of these black holes is shown to hold in the form of~\cite{AyonBeato:2009nh} 
\[   
   M = \frac{1}{1+z}T_{H}\CS_{BH}\,,
\]
which is also related to the anisotropic Cardy formula~\cite{Gonzalez:2011nz}.
On the other hand, analytic solutions for scalar hairy black holes in three dimensions have also been found in~\cite{Ayon-Beato:2015jga} and shown to satisfy the same Smarr relation with the above form. Though hairy black holes in this case seem to belong to different sector from the non-hairy black hole solutions, which have  different ground states~\cite{Correa:2010hf}, it has been shown that  they satisfy the same form of the first law of black holes and the Smarr relation.

 In this paper, we would like to explore some generic features of scalar hairy Lifshitz black holes in three dimensions and the Lifshitz planar black holes in higher than three dimensions. We show the existence of the scaling symmetry in the reduced action of new massive gravity(NMG)~\cite{Bergshoeff:2009hq} and a specific Einstein-Maxwell-dilaton(EMD)  gravity, with additional scalar fields. By using the radially conserved charge associated with the scaling symmetry, the Smarr relation of hairy black holes is derived generally. To identify the Smarr relation correctly, the mass of  scalar hairy Lifshitz black holes needs to be obtained consistently. We use the quasi-local Abbott-Deser-Tekin(ADT) formalism to identify the mass of scalar hairy Lifshitz black holes, some of which were studied in~\cite{Gim:2014nba,Liu:2014dva,Ayon-Beato:2015jga}.
 We study explicit examples of various hairy and non-hairy  black holes and confirm that our results hold in all those cases. It is straightforward to perform the  perturbative analysis on scalar hairy Lifshitz black holes to check our results.\footnote{While preparing our manuscript, we received \cite{Liu:2015tqa},  which studies asymptotic AdS case and overlaps partially  with our results.}

\section{Scalar hairy planar black holes : Set-up}
In this section, we provide our set-up and conventions with brief reviews on  some basic features of our scalar hairy planar black holes. We would like to focus on the solutions in which all the fields depend only on the radial coordinate, $r$.  In general, we can take the metric ansatz of $D$-dimensional static planar black holes with a scalar hair as
\begin{align}   \label{Ansatz}
ds^{2} = - e^{2A(r)}f(r)dt^{2} + \frac{dr^{2}}{f(r)}  + r^{2} d\Sigma^{2}_{D-2}\,, \qquad \varphi = \varphi(r)\,,
\end{align}
where $\Sigma_{D-2}$ denotes $D-2$ dimensional flat space. The gauge fields may be included and will be considered in the later section.
For instance, the asymptotic form of  the hairy Lifshitz black holes is given by 
\begin{equation} \label{Asymp}
f(r) = r^{2} + \CO(r)\,, \qquad e^{A(r)} = r^{z-1}\left[1+ \CO\Big(\frac{1}{r}\Big)\right]\,, 
\end{equation}
and the asymptotic form  of a scalar field with mass $m_\f$ can be taken as 
\begin{equation} \label{}
\varphi(r) =  \frac{\varphi_{+}}{r^{\alpha_+}} +   \frac{\varphi_{-}}{r^{\alpha_-}} + \cdots \,,
\end{equation}
where
\[   
\alpha_\pm =  \frac{D+z-2}{2} \mp \frac{1}{2}\sqrt{(D+z-2)^{2} + 4m^{2}_{\varphi}}\,.
\]
The asymptotic form includes the case of hairy AdS black holes when  $z=1$. How to choose $\varphi_{\pm}$ corresponds to  boundary conditions of scalar fields and the choice affects the role of scalar hairs in the black hole thermodynamics  and consequently their dual interpretation~\cite{Hertog:2004dr,Hertog:2004ns}.
 
Since we are considering non-rotating  scalar hairy planar  black holes, the near horizon geometry would be the product of 2-dimensional Rindler space and $(D-2)$-dimensional flat space $ \Sigma_{D-2}$. Even though the horizon radius $r_{H}$ is the unique parameter parametrizing the horizon,  the Bekenstein-Hawking-Wald entropy of planar black holes, $\CS_{BHW}$, could be a function of the horizon radius $r_{H}$ and also of the values of matter fields at the horizon. 

Regardless of the asymptotic geometry, one can take the near horizon geometry of scalar hairy black holes  as determined solely by the horizon radius as follows.
By assuming that a scalar field does not form  extremal black holes, 
the near horizon geometry of scalar hairy black holes may read as
\begin{equation} \label{}
ds^{2}_{NH} = - e^{2A(r_{H})} f'(r_{H})(r-r_{H}) dt^{2} + \frac{dr^{2}}{f'(r_{H})(r-r_{H})}  + r^{2}_{H}d\Sigma^{2}_{D-2}\,,
\end{equation}
and  the scalar field on the above geometry takes a horizon value $\varphi = \varphi(r_{H})$.  Using the change of variable $\rho^{2}=4(r-r_{H})/f'(r_{H})$, the near horizon geometry can be identified with the Rindler space as
\begin{equation} \label{}
ds^{2}_{NH} = - \frac{1}{4} e^{2A(r_{H})}f'^{2}(r_{H})\rho^{2} dt^{2} +d\rho^{2} + r^{2}_{H}d\Sigma^{2}_{D-2}\,,
\end{equation}
from which one can read the Hawking temperature as 
\[   
T_{H} = \frac{1}{4\pi} e^{A(r_{H})}f'(r_{H})\,.
\]
Recall that the Bekenstein-Hawking-Wald  entropy of black holes  would be given by the null Killing vector $\xi_{H}$ on the horizon $\CH$ as~\cite{Wald:1993nt,Iyer:1994ys}
\beq
\frac{\kappa}{2\pi} \CS_{BHW} = \frac{1}{16\pi G} \int_{\CH} dx_{\mu\nu}~ \Delta K^{\mu\nu}(\xi_{H})\,,
\eeq
where $K^{\mu\nu}$ denotes the Noether potential for a diffeomorphism and $\kappa$ does the surface gravity which is related to the Hawking temperature, $T_{H} =\kappa/2\pi$. 

%
%
One of the interesting questions about  scalar hairy black holes is whether the scalar hair has its own chemical potential in the first law of black hole thermodynamics. In other words, one may wonder the simplest form of the first law of black hole thermodynamics given in the form of 
\begin{equation} \label{}
dM = T_{H}\,  d\CS_{BHW}\,,
\end{equation}
is valid even with scalar hairs. In addition, one can ask what happens in the form of the Smarr relation  in the presence of scalar hairs. It has been known that, depending on the boundary conditions for the scalar field,  the associated chemical potential exists in the first law of black hole thermodynamics~\cite{Hertog:2004bb,Lu:2013ura}.
 However, all the examples studied in this paper satisfy the above simplest form of the first law of black hole thermodynamics and the Smarr relation without a scalar chemical potential, which can be related to the existence of a certain scaling symmetry.

Now, we would like to introduce briefly the quasi-local ADT formalism of conserved charges~\cite{Kim:2013zha,Kim:2013cor,Hyun:2014kfa,Hyun:2014sha,Hyun:2014nma}. By uplifting the on-shell background to the off-shell one in the conventional ADT formalism~\cite{Abbott:1981ff,Abbott:1982jh,Deser:2002rt}, one can match the ADT potential $Q^{\mu\nu}_{ADT}$ to the Noether potential $K^{\mu\nu}$ as
\begin{equation} \label{}
2\sqrt{-g}\, Q^{\mu\nu}_{ADT} ( \xi\,,\delta\Psi\,;\Psi\,)= \delta K^{\mu\nu}(\xi\,;\,\Psi) -K^{\mu\nu}(\delta\xi\,;\,\Psi\,) -2\xi^{[\mu}\Theta^{\nu]}(\delta \Psi\,;\,\Psi\,)\,,
\end{equation}
where $\xi$ denotes the Killing vector for a conserved charge and $\Psi$ does the relevant fields in the Lagrangian. The infinitesimal mass (density) of planar black holes in our interest is given by
\begin{equation} \label{}
\delta Q_{ADT}(\xi) = \frac{1}{8\pi G}\int_{r\rightarrow\infty}d^{D-2}x_{\mu\nu}\sqrt{-g}\, Q^{\mu\nu}_{ADT}(\xi)\,.
\end{equation}
One of the consequences of this expression when $\delta\xi^\mu=0$ is that conserved charges   for a exact Killing vector $\xi$ from the quasi-local ADT formalism are completely consistent with those from the covariant phase space. As is known in the covariant phase space formalism, the infinitesimal expression of conserved charges can be integrated along the parameter path in the solution space and becomes finite expression when the the infinitesimal one satisfies a certain integrability condition. 
In the following section, we will assume the existence of a certain on-shell scaling property in solutions.  This corresponds to a kind of an integrability in the parameter space, which  leads to a finite mass expression of static hairy planar black holes in terms of metric functions $A, f$ and the scalar field $\varphi$.

\section{Scalar hairy Lifshitz black holes  in NMG}

In this section we consider a specific higher curvature gravity in three dimensions called as NMG,  which is ghost-free and leads to a parity even massive graviton.  Furthermore, it is known to be consistent with a holographic c-theorem~\cite{Sinha:2010ai}  and  to allow various kinds of black holes in the analytic forms.  These  analytic  black hole solutions known in NMG include BTZ black holes~\cite{Banados:1992wn}, warped AdS black holes~\cite{Clement:2009gq} and new type black holes~\cite{Bergshoeff:2009aq}.  Another interesting class of analytic black hole solutions in NMG includes Lifshitz black holes which may give some interesting applications in condensed matter physics. We show that the scalar hairy Lifshitz black holes satisfy the same form of the first law of black hole thermodynamics and the Smarr relation of the dynamical exponent $z$.

\subsection{NMG coupled with a scalar field}
The action of  NMG with an additional massive scalar field  is given by
\begin{align} \label{NMG}
 I[g,\varphi]  = \frac{1}{16\pi G}\int d^3 x~\sqrt{-g} \Big[ \eta\Big(  \CL_{EH} +\frac{1}{m^2}\CL_K \Big)  + \CL_{\varphi} \Big]\,,
\end{align}
where  $\eta$ denotes signature, $\eta= \pm1$, and the metric and matter Lagrangians are taken as 
\begin{align}
& \CL_{EH} = R - 2\L\,, \qquad \CL_K=   R_{\mu\nu}R^{\mu\nu} - \frac{3}{8} R^2  \,,\quad \\
& \CL_{\varphi}  = - \frac{1}{2}\p_{\mu}\varphi\, \p^{\mu}\varphi - \frac{\alpha}{2} R \varphi^2 -V(\varphi) \,. 
\end{align}
We have introduced $\eta$ in order to take care of all the possible sign choices of the above three-dimensional actions and have abused the notation such as the parameter $m^{2}$ may take negative values.

The equations of motion(EOM) for the metric are given by
\begin{align}
\eta \Big(G_{\m\n} + \Lambda g_{\m\n} +  \frac{1}{2m^2}\S_{\m\n}\Big)  = T^{\varphi}_{\m\n}  \,,
\end{align}
where
\begin{align}
G_{\m\n} =&~ R_{\m\n} - \frac{1}{2}R g_{\m\n}  \,,\\
\S_{\m\n} =&~ 2\Box R_{\m\n} - \frac{1}{2} \Big( \nabla_{\m}\nabla_{\n} + g_{\m\n}\Box - \frac{9}{2}R_{\m\n} \Big)R -8 R_{\m}\,^{\r}R_{\n\r} +g_{\m\n}\Big(3R_{\r\s}R^{\r\s} - \frac{13}{8}R^2\Big) \,,\\
T_{\m\n}^{\varphi} =&~ \frac{1}{2} \partial_{\m}\varphi \,\partial_{\n}\varphi -\frac{1}{2} g_{\m\n}\, \CL_{\f} +\frac{\alpha}{2} \Big( R_{\m\n} -\nabla_{\m}\nabla_{\n} +g_{\m\n}\Box \Big)\varphi^2 \,,
\end{align}
and EOM for the scalar field $\varphi$ is given by
\begin{align}
\Box\varphi -\alpha R\varphi=\frac{\partial V(\varphi)}{\partial \varphi}\,.
\end{align}

These EOM allow the AdS space as a solution and the radius, $L$, of the AdS space is determined by the Lagrangian parameters through the relation  
\[   
-\Lambda L^{2} = 1- \frac{1}{4m^{2}L^{2}} \,.
\]
In the following, we have set the radius $L$, of the AdS space as unity. 
Then, the parameter $m^{2}$ in the Lagrangian is related to the cosmological constant by EOM as $1/m^2 =  4(1 + \Lambda)$.

The action of the NMG admits the Lifshitz space, of which metric can be written as
\begin{align}
ds^2 = - \Big(\frac{r}{L_z}\Big)^{2z}dt^2 + L_z^{2}\frac{dr^2}{r^2} +r^2 d\th^2\,.
\end{align}
It turns out that the radius $L_z$ is related to the Lagrangian parameters as
\[   
-L_z^{2}\L= 1 + \frac{(z^2+z-1)(z^2-3z+1)}{4 m^2 L_z^2 }, \quad  m^2 L_z^2 = \frac{1}{2}( z^2 -3z +1 )\,.
\]
By taking the Lifshitz radius $L_z$ as a unity, one obtains the relations among Lagrangian parameters as $\L = -\frac{1}{2}( z^2+z +1 )$ and $m^2 = \frac{1}{2}( z^2 -3z +1 )$.

\subsection{Reduced action of NMG}
For the given ansatz, one can rewrite the original Lagrangian of NMG in terms of fields $A$, $f$ and $\varphi$, which would become the, so-called, reduced action of NMG. The reduced action of NMG coupled with a scalar field can be written as 
\begin{align}   \label{}
I_{red} [A,f,\f]  &= \frac{1}{16\pi G}\int d^{3}x ~ L_{\varphi NMG}\,, \qquad  L_{\varphi NMG} \equiv \eta \Big(L_{EH} + \frac{1}{m^{2}}L_{K}\Big)  +L_{\f}\,,  
\end{align}
where   $L_{EH}$, $L_{K}$, and $L_{\f}$ are given,  respectively, by
\begin{align}   \label{}
L_{EH} & = -e^{A}\Big[ 2r\L + 2(A' + r A'^{2}+rA'' )f + (2 + 3rA')f' + rf'' \Big]\,,  \qquad {}'\equiv\frac{d}{dr} \,,\nn \\
L_{K}   &= \frac{e^{A}}{8r}\Big[ \Big( 2( -A'+rA'^{2} +rA'')f + 3rA'f' + rf'' \Big)^{2} -  4rf'\Big( 2(A'^{2}+A'')f +3A'f'+ f''  \Big) \Big]\,,  \nn \\
L_{\f}&= -\frac{e^{A}}{2}\Big[  rf\f'^{2} + 2rV(\f) - \alpha\f^2 \Big( 2(A' + r A'^{2}+rA'' )f + (2 + 3rA')f' + rf'' \Big)\Big]\,. \nn
\end{align}
This reduced action contains four derivative terms and the analysis becomes complicated in this form.  
To simplify the analysis, one can introduce auxiliary fields taking the role of reducing the derivative orders. 
Note that the above reduced action can be put in the following form 
\begin{align}   \label{}
 L_{\varphi NMG}
 &= - e^{A}\bigg[ \eta\Big( 2r\L + f'  - 2f\lambda' - f'\lambda\Big) +\frac{1}{2}rf\f'^{2} + rV(\f) - \frac{\a}{2}e^{-A}(Z' + e^{A}f')\f^2 \bigg]  \nn  \\
&\qquad \qquad  + \frac{\eta e^{-A}}{8m^{2}r}\bigg[ 8e^{A}\lambda Z + ( Z'-e^{A}f')^{2} -\frac{4}{r}ZZ' + \frac{4}{r^{2}}Z^{2}\bigg]   - \eta \Big(Z+\frac{2}{m^{2}}e^{A}f\lambda\Big)' \,, 
\end{align}
where  $\lambda$ is introduced as a  Lagrange multiplier for our convenience and then we have used 
\begin{equation} \label{}
-\lambda(2e^{A}f)' = -( 2e^{A}f \lambda)'  + 2e^{A}f \lambda'\,.
\end{equation}

It is straightforward to obtain EOM by taking variations with respect to $A, f, Z$ and $\lambda$. Note that the last term of the above reduced action is a total derivative and so it would be dropped in the following. One can see that  $e^{A}$ is  another  Lagrange multiplier. As a result, the dynamical EOM come from the variations with respect to $f$ and $Z$. 
Concretely, EOM by the variation of $\lambda,  Z$ and $e^{A}$ are given, respectively, by
\begin{align}   \label{}
Z &= (2re^{A}f)' - 2e^{A}f-re^{A}f' = re^{A}f' +2r(e^{A})'f,   \\
\lambda & = \frac{\eta\alpha}{2} r (\varphi^{2})' +\frac{r}{4m^{2}}\bigg[\frac{e^{-A}}{r}(Z'-e^{A}f')\bigg]' - \frac{Z}{2m^{2}r}(e^{-A})'\,,    \label{lambda}   \\
0 &= - \frac{r}{2}f\f'^{2} - rV(\f)  +\frac{\a}{2}f'\f^2  \nn  \\
& \quad + \eta \bigg[ -f'  -2r\Lambda + 2 f \lambda' +  f' \lambda  - \frac{e^{-2A}}{8m^{2}r}\Big(Z'-\frac{2}{r}Z\Big)^{2} + \frac{f'^{2}}{8m^{2}r} \bigg] \label{Avar}\,. 
\end{align}
The dynamical EOM from the variations of $f$ and $\varphi$ are given, respectively, by
\begin{align}   \label{}
0&= -\frac{r}{2}e^{A}\f'^{2}- \frac{\alpha}{2}(e^A\f^2)' + \eta\bigg[(e^{A})'  + \frac{1}{4m^{2}}\Big\{ \frac{1}{r}(Z'-e^{A}f')\Big\}'  + e^{A}\lambda' -  (e^{A})' \lambda \bigg]\,, \\
0 &= (r e^A f\f')' + e^A ( \a f'\f- r \partial_{\varphi} V) \,.
\end{align}
One can check that these EOM are consistent with  EOM from the original Lagrangian given in the appendix. Note also that the above EOM reproduce those of Einstein gravity when $m^{2}\rightarrow \infty$.

This reduced action is invariant under the rescaling of $r$, which can be realized, infinitesimally, in terms of  field variations as
\begin{align}   \label{}
\delta_{\sigma}f = \sigma (2f-rf')\,, \qquad \delta_{\s} e^{A} = \sigma(-2-rA')e^{A} \,, \qquad \delta_{\s} \f & = -\sigma r \f'\,. 
\end{align}
Generically, any field $\Psi$ of the scaling weight $w$  in the reduced action  transforms as 
\begin{equation} \label{}
\delta_{\s} \Psi = \sigma (w\Psi-r\Psi')\,.
\end{equation}
It is straightforward to check that the reduced action transforms under the above scaling symmetry as
\begin{equation} \label{}
\delta_{\sigma} I_{red} =   \frac{1}{16\pi G}\int d^3 x\, S'\,,
\end{equation}
where the total derivative term $S$ is given by
\begin{equation} \label{}
S = - r \Big[ \eta \Big( L_{EH}+\frac{1}{m^{2}}L_{K}\Big) +L_{\f}\Big]\,.
\end{equation}
Under a generic variation, the reduced action transforms as 
\begin{equation} \label{}
\delta I_{red} =  \frac{1}{16\pi G} \int d^3 x\, \Big[ \CE_{f}\delta f +\CE_{Z}\delta Z+ \CE_{A}\delta A +\CE_{\lambda}\delta\lambda+ \CE_{\f}\delta \f + \Theta'(\delta\Psi)\Big]\,,
\end{equation}
where $\CE_{\Psi}$ denotes the Euler-Lagrange expression of the field $\Psi$ and $\Theta$ denotes the surface term under a generic variation. 
Since derivative terms of $A$ are absent in the reduced action, the total surface term $\Theta$ is composed of those from $\delta f, \delta Z, \delta \lambda,$ and $\delta\f$ as follows:
\begin{equation} \label{}
\Theta(\delta \Psi) = \Theta(\delta f) + \Theta(\delta Z)  + \Theta(\delta \lambda)+\Theta(\delta \f)\,,
\end{equation}
where each surface term is given by 
\begin{align}   \label{}
\Theta(\delta f) &= e^{A}\bigg[ \eta\Big\{ -1 +  \lambda  - \frac{1}{4m^{2}r} (Z'- e^{A}f') \Big\}+ \frac{\a}{2}e^A \f^2 \bigg]\delta f\,, \nn \\
\Theta(\delta Z) &= \bigg[\frac{\a}{2}\f^2+\frac{\eta e^{-A}}{4m^{2}r} \Big(Z'- e^{A}f' -\frac{2}{r}Z\Big)\bigg]\delta Z\,, \nn \\
\Theta(\delta \lambda ) & = 2\eta e^{A}f \delta \lambda\,, \nn \\
\Theta(\delta \f) & = -re^{A}f\f' \delta \f\,. \nn
\end{align}

The Noether charge  for the scaling symmetry\footnote{Here, we take  $\sigma =1$ for the parameter independent charge expression.}, which remains constant along $r$, is given by 
\begin{equation} \label{NoetherC}
 C(r)=\frac{1}{8G} \Big[\Theta(\delta_{\sigma}\Psi) -S \Big]\equiv C \,.
\end{equation}
As a result, the conserved charge  for the scaling symmetry in NMG is given by
\begin{align}   \label{}
8G\, C(r) 
&= e^{A}\Big[-\eta( 2 f+ 2 r^{2}\Lambda) - r^{2}V(\f) + \frac{1}{2}r^{2}f\f'^{2} + \a f\f^2\Big]  + \eta\lambda (Z+2 e^{A} f ) \nn \\
&~~~ - \frac{\eta e^{-A}}{8m^{2}}\Big[(Z'-e^{A}f')^{2} - \frac{4}{r^{2}}Z^{2}\Big] - \frac{\eta f}{2m^{2}r}(Z'-e^{A}f')\,.
\end{align}
We use EOM from the $e^{A}$-field variation in Eq.~(\ref{Avar}) to express the charge $C$ in the form, which is independent of the scalar potential $V(\varphi)$, just like the Einstein gravity case~\cite{Banados:2005hm}, as follows
\begin{align}   \label{Noether}
 8G\, \eta C(r) =\,& e^{A}\Big[ -1+\lambda - \frac{e^{-A}}{4m^{2}r}(Z'-e^{A}f')  +  \eta \frac{\a}{2}\varphi^{2}  \Big](2f-rf') + \lambda Z \nn \\
&   + e^{A}\Big[- 2rf\lambda'   + \eta\,  r^{2} f\varphi'^{2} \Big]  + \frac{e^{-A}}{2m^{2}r^{2}}Z(2Z-rZ')  \,.
\end{align}

Let us clarify the meaning of the conserved charge $C$ by comparing it with the well-known physical quantities of black holes. The only conserved physical quantity which is defined at the event horizon of the static black hole is the Bekenstein-Hawking-Wald entropy~\cite{Wald:1993nt}.
By using the covariant phase space method or, equivalently,  the quasi-local ADT formalism,  one can obtain the black hole entropy  as
\begin{equation} \label{}
\frac{\kappa}{2\pi}\CS_{BHW}  = \frac{1}{8\pi G}\int ds \int_{\CH} dx_{\mu\nu}\sqrt{-g}Q^{\mu\nu}_{ADT}(\xi_{H}\,;\,\Psi \,|\, s)  = \frac{1}{16\pi G}\int_{\CH} dx_{\mu\nu}\Delta K^{\mu\nu}(\xi_{H}\,;\,\Psi) \,,
\end{equation}
where the relevant Noether potential, for the null Killing vector $\xi_H$, is given by
\begin{equation} \label{}
K^{rt}(\xi_H\,;\,\Psi)  =  \bigg[ \eta\bigg(  Z + 2e^{A}f\lambda - \frac{e^{-A}}{4m^{2}r}Z(Z'-e^{A}f') + \frac{e^{-A}}{2m^{2}r^{2}}Z^{2} \bigg)-\frac{\a}{2}Z\f^2 \bigg]_{r=r_H} \,.
\end{equation}

By using the fact  that $f(r_{H}) =0$  at  the horizon $r=r_H$,
the conserved charge $C$ in Eq.~(\ref{Noether}) can be expressed as
\begin{equation} \label{}
 C = \frac{\eta}{8G} e^{A(r_{H})}  f'(r_{H}) \Big[ r_{H} \Big(1 - \eta \frac{\alpha}{2}\varphi^{2}\Big)+ \frac{1}{4m^{2}}\Big(2f'(r_{H}) -3r_{H}A'(r_{H})f'(r_{H})-r_{H}f''(r_{H})\Big)\Big]\,.
\end{equation}
It is straightforward to check that
\begin{equation} \label{Centropy}
C=\frac{\kappa}{2\pi}\CS_{BHW} \,,
\end{equation}
where $\kappa$ is the usual surface gravity and so the Hawking temperature is given by 
\[    T_{H}=\frac{\kappa}{2\pi }=  \frac{1}{4\pi} e^{A(r_{H})}f'(r_{H})\,.
\]

Now, let us turn to the conserved physical quantity at the asymptotic infinity. In our case at hand, it is nothing but the total mass. In order to determine the total mass, we apply
 the quasi-local ADT formalism, which was explained shortly in section 2, to NMG in our ansatz.  The ADT potential for the Killing vector $\xi_{T}=\frac{\partial}{\partial t}$ in terms of metric fields $A, f$ and the scalar field $\varphi$ is given in the appendix~(\ref{Qadt}).
This ADT potential  can be rewritten, in terms of the auxiliary fields,  as 
\begin{align}   \label{potential}
2\sqrt{-g}Q^{rt}_{ADT}(\xi_{T}\,;\, \delta g) 
=&  \eta\bigg[ e^{A}\Big\{ -1 + \lambda - \frac{e^{-A}}{4m^{2}r}(Z'-e^{A}f') + \eta \frac{\a}{2} e^{A}\varphi^{2} \Big\} \delta f  + 2fe^{A}\, \delta \lambda     \nn \\
&~ - \frac{e^{-A}}{4m^{2}r}Z\, \delta Z'   +\frac{Z}{4m^{2}r} \delta f'  + \frac{e^{-A}}{2m^{2}r^{2}}\, Z\delta Z     -\frac{e^{-A}}{4m^{2}r^{2}}Z(2Z - rZ')\delta A \bigg]\,, \nn  \\  \nn \\
2\sqrt{-g}Q^{rt}_{ADT}(\xi_{T}\,;\, \delta \varphi)
=&  -  (r e^A f \, \varphi'   + \alpha Z \varphi )\delta \varphi\,. 
\end{align}
One may note that the derivatives of $\delta A$ are hidden in the variations of  auxiliary fields $\lambda$, $Z$ and $Z'$.
The infinitesimal quasi-local ADT  mass expression of black holes is given by
\begin{equation} \label{ADTmass}
\delta M_{ADT}  = \frac{1}{8\pi G} \int_{r\rightarrow\infty} dx_{\mu\nu} \sqrt{-g}\Big( Q^{\mu\nu}_{ADT}(\xi_{T}\,;\, \delta g) +Q^{\mu\nu}_{ADT}(\xi_{T}\,;\, \delta \varphi)\Big)\,. 
\end{equation}

In order to get the finite mass expression, we need to integrate this infinitesimal expression $\delta M_{ADT} $ along the one-parameter 
path in the solution space. In the case at hand we choose specific one-parameter path as follows.  There exists an $r$-coordinate scaling transformation of metric and scalar fields which  preserves EOM. By supplementing an appropriate coordinate transformation $\delta_{Diff}$, this gives the same form of the metric and scalar fields while changing the values of parameters in solutions~\cite{Horowitz:1999jd}. This gives a natural one-parameter  path in the solution space.   This `on-shell'  scaling of $r$-coordinate in black hole solutions can be realized as field variations  of $f, e^{A}, \lambda, Z$ and $\varphi$. Explicitly, let us introduce 
\begin{align}   \label{}
\hat{\delta}_{\sigma}f &= 2f - rf'\,, \qquad \qquad  \hat{\delta}_{\sigma}\lambda  = - r\lambda'\,, \qquad \qquad  \hat{\delta}_{\sigma}\varphi = -r\varphi' \,, \\ 
\hat{\delta}_{\sigma}e^{A} &= \gamma e^{A}-r(e^{A})'\,, \qquad \hat{\delta}_{\sigma}Z = (\gamma+2)Z- rZ'\,. \nn 
\end{align}
One may note that  EOM remain the same under the above variations with an arbitrary $\gamma$, while these become the symmetry transformations of the reduced action only when $\gamma= -2$.  If  the value of  $\gamma$ is chosen as  $\gamma=z-1$,  the  forms of the metric and scalar fields remain the same while changing the values of parameters.  Since the ADT mass expression in Eq.~(\ref{ADTmass}) is invariant under the diffeomorphism, 
the variation of the mass $M$ along the path corresponding the scaling transformation with ${\gamma=z-1}$ supplemented by the coordinate transformation, {\it i.e.}, $\delta=\hat{\delta}_\sigma+\delta_{Diff}$, is given by
\begin{equation} \label{}
\delta M_{ADT} =\hat{\delta}_{\sigma} M_{ADT} = \sigma(1+z) M_{ADT}\,.
\end{equation}

A simple computation by using the expression of $\lambda$ given in Eq.~(\ref{lambda}) reveals  that the ADT potential in Eq.~(\ref{potential}) turns out to have the same expression as the Noether charge in Eq.~(\ref{Noether}) as 
\begin{align}   \label{}
\frac{1}{4G}\left[\sqrt{-g}Q^{rt}_{ADT}(\xi_{T}\,;\, \hat{\delta}_{\sigma} g)+\sqrt{-g}Q^{rt}_{ADT}(\xi_{T}\,;\, \hat{\delta}_{\sigma} \varphi)\right] 
= \sigma C\,.
\end{align}
As a result, we have the relation: 
\begin{equation} \label{}
\sigma C= \sigma C(r\rightarrow \infty) = \hat{\delta}_{\sigma}M_{ADT} = \sigma (1+z)M_{ADT}\,.
\end{equation}
By recalling that  $C(r)$ is invariant along the radial direction $r$ and the Eq.~(\ref{Centropy}), we arrive at the generic result on the Smarr relation of hairy Lifshitz black holes as 
\begin{equation} \label{}
(1+z)M = T_{H}\CS_{BHW}\,.
\end{equation}
This is one of our main results for hairy Lifshitz black holes in three dimensions. 

\subsection{Examples}
In this section, we check our results for various examples explicitly. 
In three dimensional gravity it is straightforward  to obtain  the  conserved charge $C$. First, let us consider non-hairy black holes in NMG. At the horizon for all these black holes, it is trivial to confirm that the conserved charge is related to the Bekenstein-Hawking-Wald entropy as in Eq.~(\ref{Centropy}). At the asymptotic infinity, as shown in below, one can see that the conserved charge is proportional to the mass of the black hole.  

The BTZ black holes exist  for any value of the parameter $m^{2}$. The non-rotating BTZ black holes  correspond to the metric functions of the form $e^{A}=1$ and $f(r)= r^{2} - a$. One can check that the conserved charge is related to the mass~\cite{Bergshoeff:2009aq,Oliva:2009ip}
of the non-rotating BTZ black hole as 
\begin{equation} \label{}
C= \eta \frac{a}{4G} \Big[1+ \frac{1}{2m^{2}}\Big] =2M_{ADT}\,.
\end{equation}

The
new type black holes~\cite{Bergshoeff:2009aq,Oliva:2009ip,Nam:2010ub} have been known to exist
when the parameter takes the value $m^{2}=1/2$.  The non-rotating black holes are given by the metric functions of the form   $e^{A}=1$ and $f(r) = r^{2} + br +c$. In these new type black holes, we again find the relation between the conserved charge and the mass as
\begin{equation} \label{}
 C = \frac{\eta}{8G} (b^{2} - 4c)=2M_{ADT}\,. 
\end{equation}

At the point $m^{2}= 1/2$, the  Lifshitz black holes are also found with the anisotropic scaling $z=3$~\cite{AyonBeato:2009nh}. They are given by taking $e^{A} = r^{z-1}$ and $f = r^{2} -a$, in which one can see that 
\begin{equation} \label{}
C = - \eta \frac{a^{2}}{G}=(1+z)M_{ADT}\,.
\end{equation}

Now we turn to the examples of scalar hairy black holes.
The scalar hairy AdS black holes in Einstein gravity was found in~\cite{Henneaux:2002wm} and the relevant asymptotic forms of the metric functions and the scalar field are given by
\begin{align}   \label{}
e^{A}& = 1 - \frac{2B}{r} + \frac{6B^{2}}{r^{2}} + \cdots\,, 
  \nn \\
 f &= r^{2} +4B r -3(1+\nu)B^{2} + \cdots\,, \\
 \varphi  &=  \frac{4\sqrt{B}}{r^{1/2}} - \frac{8B^{3/2}}{3r^{3/2}} + \cdots\,. \nn 
\end{align}
The charge $C$ is given by
\begin{equation} \label{}
C = \frac{3}{4G}B^{2}(1+\nu) = 2M_{ADT}\,. 
\end{equation}

For hairy Lifshitz black holes in NMG given in~\cite{Ayon-Beato:2015jga}, one  can take 
\begin{equation} \label{}
e^{A} = r^{z-1}\,, \qquad f = r^{2} - \frac{a}{r^{\frac{z-3}{2}}}\,, \qquad \varphi = \sqrt{\frac{(z - 3) (9 z^2 - 12 z + 11) a}{(z - 1) (z^2 - 3 z + 1)}}  \frac{1}{r^{\frac{z+1}{4}}}~.
\end{equation}
Therefore we find that
\begin{equation} \label{}
\, C = -\frac{\eta }{8G} \frac{a^2 (1 + z)^3 (-5 + 3 z)}{16 (-1 + z) (1 - 3 z + z^2)}=(1+z)M_{ADT}\,. 
\end{equation}

In all these black holes, we confirm the Smarr relation through the conserved charge as
\begin{equation} \label{}
C =T_{H}\CS_{BHW}=  (1+z)M_{ADT}\,.
\end{equation}
%

\section{Lifshitz planar black hole solutions in higher dimensions}
%
%

In this section, we consider the specific EMD gravity ~\cite{Taylor:2008tg,Tarrio:2011de} which admits Lifshitz planar black hole solutions  in higher dimensions.
The model allows the asymptotically Lifshitz structure with anisotropic scale invariance when the gauge and dilaton fields have nontrivial asymptotic profiles.
The action is given by
\begin{equation}
	I[g,\phi,\CA] = \frac{1}{16\pi G} \int d^{D}x \sqrt{-g} \Big( \CL_{EH} +\CL_{\phi \CA}   \Big) \,, \qquad \CL_{\phi \CA}\equiv-\frac{1}{2}\big(\partial\phi \big)^2 - \frac{1}{4}e^{\lambda\phi }\CF^{2}\,,
\end{equation}
where, to support asymptotic Lifshitz geometry in $D$ dimensions, the parameters are chosen as
\begin{align}
 \l =-\sqrt{2 \frac{D-2}{z-1}} \,, \qquad 
\L = -\frac{(D+z-2)(D+z-3)}{2}\,. \nn
\end{align}
The AdS background can be obtained by taking the formal limit $z\rightarrow 1$ under which the gauge field is decoupled.   To obtain the asymptotic Lifshitz geometry given by Eqs.~(\ref{Ansatz}) and (\ref{Asymp}), the asymptotic forms of the dilaton and the gauge fields should be taken as
\begin{align}   \label{EMDasym}
	e^{\phi} =&~ \,  \mu \,r^{\sqrt{2(D-2)(z-1)}} \left[1+\CO\Big(\frac{1}{r}\Big) \right] \,,\\
	\CA=&~   \sqrt{ \frac{2(z-1)}{D+z-2} }\, \mu^{-\frac{ \l }{2} }\, r^{D+z-2} \left[1+\CO\Big(\frac{1}{r}\Big) \right]\,dt \,, \qquad \CF=d\CA\,.
\end{align}
One may notice that we need to introduce the dimensionful parameter $\mu$ which plays the role to support the asymptotic Lifshitz structure and has nothing to do with black hole geometry.
In order to engineer the scalar hair in Lifshitz black holes, we introduce an additional scalar field with the action
\begin{align}
 I_{\f} = \frac{1}{16\pi G} \int d^{D}x\, \sqrt{-g} \left[- \frac{1}{2} \big( \partial \f \big)^{2} - V(\f) \right] \,.
\end{align}

We consider the reduced action formalism in this model, which can cover static, Lifshitz planar black holes with a scalar hair.
We take the ansatz for the metric and the additional scalar field $\f$ as given in Eq.~(\ref{Ansatz}). We also take the ansatz for the dilaton  and gauge fields as $\phi(r)$ and $\CA = a (r) dt$, respectively.
Then, the reduced action becomes
\begin{equation} \label{}
I_{red} [A,f,a,\phi,\varphi]  = \frac{1}{16\pi G}\int d^{D}x\,  \Big(L_{EH} + L_{\phi\CA} +L_{\varphi} \Big)\,,
\end{equation}
where the Lagrangian $L_{EH}$, $L_{\phi\CA}$ and $L_{\f}$ are given, respectively, by
\begin{align} \label{}
L_{EH}&= -e^{A} \bigg[  \Big( (r^{D-2})' f \Big)' + 2r^{D-2} \Lambda   \bigg] -  \bigg[ r^{D-2}  e^{A} f' +   2r^{D-2}  (e^{A})'  f\bigg]' \,,\\
L_{\phi\CA}  &= -\frac{1}{2} r^{D-2}  e^{A} f \phi'^{2} + \frac{1}{2}r^{D-2} e^{-A} e^{\lambda \phi}  a'^2 \,,\\
L_{\varphi}&=  -\frac{1}{2}r^{D-2}e^{A}   f \varphi'^{2}  -r^{D-2}e^A V(\varphi)  \,.
\end{align}
From the variations with respect to the each field, $f\,,e^A\,,a\,,\phi$ and $\f$ in the reduced action, EOM are given as follows:
\begin{align}   \label{}
0&=\Big(e^{A} (r^{D-2})' \Big)'-\frac{1}{2} r^{D-2} e^{A}\, ( \phi'^2+\f'^2) \,, \\
0 &= r^{2-D} \Big( (r^{D-2})' f \Big)' +2 \Lambda + \frac{1}{2}  f \varphi'^{2} + V + \frac{1}{2} f \phi'^{2} + \frac{1}{2} e^{-2A}  e^{ \lambda \phi} a'^{2} \,,\label{EOMA} \\
0&= ( r^{D-2} e^{-A}e^{\lambda \phi}  a' )' \,, \label{EOMgauge} \\
0&= (r^{D-2} e^A f \phi')' + \frac{1}{2}r^{D-2}e^{-A} \lambda e^{\lambda \phi }  a'^{2} \,, \label{EOMdilaton} \\
0 &= (r^{D-2} e^A f \varphi')' - r^{D-2}e^A\,  \partial_{\varphi} V \,.
\end{align}
One may note that these EOM  can also be derived from the original Lagrangian.
%

%

The reduced action is invariant under the following  infinitesimal field transformations:
\begin{align}
	\d_{\s} f &= \s (2f -r \, f') \,, \quad \d_{\s} e^{A} = \s \big( -(D-1) -r \, A' \big) e^{A} \,, \\  
	\d_{\s} \varphi &= -\s r \, \varphi' \,, \quad \d_{\s} \phi = -\s r \, \phi' \,, \quad \d_{\s} a = \s \big( -(D-2) a -r \, a' \big)\,.
\end{align}
The Noether charge $C$ for this  scaling symmetry is given by the same expression as in Eq. (\ref{NoetherC}).
The relevant total derivative term $S$ and the total surface term $\Theta$ for this model are given by
\begin{align}
S&=-r(L_{EH}+L_{\phi\CA}+L_{\varphi} )\,,\\
\Theta(\d \Psi)&=  - (r^{D-2})' e^{A}\d f  + r^{D-2} e^{-A} e^{\lambda \phi} a'  \,\d a- r^{D-2} e^{A} f \phi'  \d \phi   - r^{D-2} e^{A} f \varphi'  \d \varphi \,.
\end{align}
%
%
%

Consequently, the Noether charge, which remains constant along $r$, is given by  
\begin{align}
8G C(r)=&~ -r^{D-3} e^{A} \bigg[ (D-2)(D+1)  f  +\frac{1}{2} r e^{-2A} e^{\lambda \phi} \Big( r a'^2   + 2(D-2)  a\, a'\Big)  \nn\\
&\qquad\qquad\qquad  + 2r^2 \Lambda  - \frac{1}{2}r^2  f \phi'^{2} -  \frac{1}{2} r^2 f \varphi'^{2}+  r^2 V(\varphi) \bigg] \,.
&
\end{align}
One may note that we have two more integrals of motion. By integrating the gauge field EOM given in Eq.~(\ref{EOMgauge}), we obtain one integration constant $q$
\begin{equation} \label{q}
r^{D-2} e^{-A}e^{\lambda \phi}  a'  =\, q\,,
\end{equation}
which corresponds to the dimensionful parameter $\m$.
By plugging Eq.~(\ref{q}) to the dilaton field EOM given in Eq.~(\ref{EOMdilaton}), we obtain another integration constant $X$
\begin{align} \label{X}
r^{D-2} e^A f \phi' + \frac{1}{2}\, \lambda \,q \, a =X\,.
\end{align}
By using  Eq.~(\ref{EOMA}) and Eq.~(\ref{q}), one can rewrite the conserved charge $C$ as 
\begin{align}\label{C}
8GC=&  - r^{D-3}e^{A} \bigg[ (D-2) \Big(2f  - r f'  \Big) - r^{2}f \phi'^{2}  -   r^{2}f \f'^{2}     \bigg] -(D-2)\,q\,a \,.
\end{align}

Now, we are in the position  to compare this Noether charge with the physical quantity of the black hole at the horizon, namely the Bekenstein-Hawking-Wald entropy. The entropy density $\CS_{BHW}$ can be obtained by the similar procedure as in the section 3 and is given by
\begin{align}   \label{}
\frac{\k}{2\pi} \CS_{BHW} ~ \Sigma_{D-2}= \frac{1}{16\pi G} \int_{\CH} dx_{\mu\nu}2\nabla_{\phantom T}^{[\mu}\xi^{\nu]}_T  = \frac{1}{16\pi G}\,r_H^{D-2}e^{A(r_H)}f'(r_H)\, \Sigma_{D-2}\,.
\end{align}
%
The expression of the conserved charge at the horizon $r=r_H$ becomes
\begin{align}
C=&  \frac{D-2}{8G}   \Big[ r_H^{D-2}e^{A(r_H)} f'(r_H)  - q a(r_H)\Big]  \,.
\end{align}
Henceforth, contrary to the NMG case in the previous section,  the   charge $C$ at the horizon $r=r_{H}$ is not directly related to  $T_{H}\CS_{BHW}$. Rather,  one can  check that 
\begin{equation} \label{}
(D-2)T_{H}\CS_{BHW}  = \frac{1}{2\pi}\left( C-C_{0}\right)\,, \qquad  C_0  \equiv - \frac{D-2}{8G} q a(r_H)\,. 
\end{equation}
Since the integration constant $X$ given in Eq.~(\ref{X}) can be written in terms of the  value at the horizon as $X=-\sqrt{\frac{D-2}{2(z-1)}}\,qa(r_H)$, we can  confirm that  $C_0$ is proportional to the integration constant $X$.
Like in the previous section, we would like to compare  the charge $C$ at the  asymptotic infinity with the ADT mass expression. 
In this model, the ADT potentials for the timelike Killing vector $\xi_T$ are given by
\begin{align}   \label{ADTEMD}
2\sqrt{-g}\,Q^{rt}_{ADT}(\xi_{T}\,;\, \delta g) &=  -(D-2) r^{D-3} e^A\delta f +r^{D-2}e^{-A}  e^{\lambda \phi}a\, a' \delta A \,,\\
2\sqrt{-g}\, Q^{rt}_{ADT}(\xi_{T}\,;\, \delta {\cal A}) &= -r^{D-2} e^{-A} e^{\l \phi} a\,\delta a' \,,\\
2\sqrt{-g}\, Q^{rt}_{ADT}(\xi_{T}\,;\, \delta \phi) &= -r^{D-2} e^{-A}  \l  e^{\lambda \phi}a\,a' \delta\phi -r^{D-2} e^A f\phi' \, \delta\phi   \,,\\
2\sqrt{-g}\, Q^{rt}_{ADT}(\xi_{T}\,;\, \delta \varphi) &= -r^{D-2} e^A f \varphi' \, \delta\varphi \,.
\end{align}
Using the integral of motion in Eq.~(\ref{q}), one can see that the on-shell variation should satisfy
\begin{equation} \label{}
- r^{D-2}e^{-A} e^{\lambda \phi}  a' \delta A  +  r^{D-2}e^{-A} e^{\lambda \phi}  \delta a' +  r^{D-2}e^{-A} \lambda e^{\lambda \phi}  a' \delta\phi =\delta q\,,
\end{equation}
which leads to the infinitesimal mass formula in the quasi-local ADT formalism as
\begin{align}   \label{TVmass}
 \delta M_{ADT} &= \frac{1}{8\pi G} \int_{r\rightarrow\infty}dx_{\mu\nu}\sqrt{-g}\,Q^{\mu\nu}_{ADT}(\xi_{T}\,;\, \delta \Psi) \nn \\
&= \frac{1}{16\pi G}\Big[ -(D-2) r^{D-3} e^A\delta f  - a\,\delta q -r^{D-2} e^A f (\phi'   \, \delta\phi  +  \varphi' \, \delta\varphi) \Big]_{r\rightarrow\infty} \,.
\end{align}
%
%
 
To obtain the mass expression of these Lifshitz planar black holes, we  adopt a specific one-parameter path in the solution space, which is induced by the `on-shell' scaling transformation as before.  EOM remain the same under the following `on-shell' scaling  transformation with an arbitrary $\gamma$,
\begin{align}\label{onshell2}
	\hat{\d}_{\s} f &= \s (2f -r \, f') \,, \quad \hat{\d}_{\s} e^{A} = \s \big( \g -r \, A' \big) e^{A} \,, \nn\\  
	\hat{\d}_{\s} \varphi &= -\s r \, \varphi' \,, \quad \hat{\d}_{\s} \phi = -\s r \, \phi' \,, \quad \hat{\d}_{\s} a = \s \big( (\g+1) a -r \, a' \big)\,.
\end{align}
This becomes the symmetry of the reduced action only when $\gamma= -(D-1)$, and preserves the  forms of the metric, gauge, dilaton and scalar fields  with the appropriate rescaling of the parameters  when  $\gamma=z-1$.

However, one needs to be cautious in taking the one-parameter path when the background dimensionful parameter $\mu$ as given in Eq.~(\ref{EMDasym}) exists, as it has nothing to do with the black hole geometry. Since the `on-shell' scaling along the radial coordinate requires inevitably the change of the background dimensionful parameter $\mu$,  which does not correspond to the parameter path in the solution space for black holes,  one needs to subtract this unwanted change  from the quasi-local ADT mass expression in Eq.~(\ref{ADTmass}).  Explicitly, the one parameter path would be chosen as $\delta = \hat{\delta}_{\sigma} + \hat{\delta}_{\mu}+ \delta_{Diff}$, where $\hat{\delta}_{\mu}$ denotes the compensating term caused by the background parameter change as follows.  By looking at the asymptotic forms of the dilaton and gauge fields, one can see that the `on-shell' scaling of $\mu$ parameter is given by
\beq
\hat{\delta}_{\sigma} \mu = -\sigma \sqrt{2(D-2)(z-1)}\, \mu\,.
\eeq
This change  should be subtracted, to form a one-parameter path in the solution space, from the `on-shell' scaling transformation of  the  dilaton field and the integral constant $q$,  as\footnote{We have omitted the diffeomorphism transformation in the formulas since it does not contribute to the covariant ADT expression.} 
\beq
\delta \phi  = \hat{\delta}_{\sigma} \phi  + \hat{\delta}_{\mu} \phi \,,    \qquad \delta q= \hat{\delta}_{\sigma} q  +\hat{\delta}_{\mu}q\,,
\eeq
where
\beq\label{delmu}
\hat{\delta}_{\mu} \phi \equiv \sigma \sqrt{2(D-2)(z-1)}\,, \qquad  \hat{\delta}_{\mu} q \equiv -\sigma (D-2)q\,. 
\eeq
After all these  considerations, one can see that the infinitesimal form of the ADT mass expression becomes 
\beq
\delta M_{ADT} = \hat{\delta}_{\sigma} M_{ADT} + \hat{\delta}_{\mu}M_{ADT} = \sigma(D+z-2)M_{ADT}\,.
\eeq
By inserting  Eq.~(\ref{onshell2}) and $\hat{\d}_\s q=\s(D-2)q$ to  Eq.~(\ref{TVmass}), we compute to find
\beq
 \hat{\delta}_{\sigma} M_{ADT} = \frac{\sigma}{2\pi} C\,. 
 \eeq
From Eqs.~(\ref{TVmass}), (\ref{delmu}) and the relation between $X$ and $C_0$, one can also find 
\beq
 \hat{\delta}_{\mu}M_{ADT} = -\frac{\sigma }{16\pi G}\sqrt{2(D-2)(z-1)}\,X=-\frac{\sigma}{2\pi} C_0\,.
 \eeq
%
%
Finally, we obtain the generalized Smarr relation as\footnote{We have been informed~\cite{Bertoldi:2009dt,Bertoldi:2010ca,Bertoldi:2011zr} that the same relation was obtained, through a similar scaling symmetry and  the perturbative analysis, for the non-hairy Lifshitz planar black holes.}
\beq 
\frac{1}{2\pi}(C-C_0) = (D+z-2)M_{ADT}  = (D-2) T_{H}\CS_{BHW}\,.
\eeq
These results are completely consistent with those  in~\cite{Liu:2014dva,Brenna:2015pqa} when  the scalar hair is turned off.
Note that the gauge field charge does not appear in the Smarr relation, nor in the first law of black hole thermodynamics, since it play the role supporting the asymptotic Lifshitz structure. On the contrary, the charges and potentials of usual gauge fields, which fall off asymptotically,  appear in the Smarr relation as well as in the first law. 

Now, we check explicitly our results for the non-hairy black hole solutions given in~\cite{Taylor:2008tg,Tarrio:2011de}. The metic functions in our ansatz are given by
\begin{align}
e^{A(r)} = r^{z-1} \,,\quad
f(r) = r^{2} \bigg( 1-\frac{m}{r^{D+z -2}} 
 \bigg) \,,
 \end{align}
 while the dilaton and  gauge fields are
 \begin{align}
	e^{\phi} = \,  \mu \,r^{\sqrt{2(D-2)(z-1)}} \,,\quad	\CA=   \sqrt{ \frac{2(z-1)}{D+z-2} }\, \mu^{-\frac{ \l }{2} }\, r^{D+z-2} \,dt \,.
\end{align}
The ADT mass,  entropy density and  temperature of these black holes are given by 
\[
M_{ADT} = \frac{(D-2)m}{16 \pi G},\qquad \CS_{BHW}= \frac{1}{4G}m^{  \frac{D-2}{D+z-2}}, \qquad T_{H} = \frac{D+z-2}{4\pi} m^{\frac{z}{D+z-2}}.
\]
Since the Noether charge $C$ and the integration constant $C_0$ can computed as
\[
C=(D-2)(D-z)\frac{m}{8G}\,,\quad C_{0} = -2 (D-2)(z-1)\frac{m}{8G}\,,
\]
we  confirm our results in this example.

\section{Conclusion}  
The advent of the AdS/CFT correspondence and its cousins has led to much interest on  black holes with the non-flat asymptotic structure. Along this correspondence, various scalar hairy black holes in the AdS space  have been found in the analytic form, which may be  contrasted  to the sacred folklore known as the no-hair theorem. Though the analytic solutions of scalar hairy black holes have been found, their stability is not manifest and so no-hair theorem still gives us some guidelines and plays some role in understanding black hole physics. Since the understanding  of the true nature of no-hair theorem or the hairy black hole with the non-flat asymptotic geometry  is incomplete,  it would be better to have  a model-independent results on hairy black holes. 

In this paper, we have explored the scaling symmetry of the reduced action for  static planar black holes and found that a generalized Smarr relation can be derived in a very general way.  Particularly, we have derived the Smarr relation for scalar hairy Lifshitz black holes, which cover the AdS space as a special case. Though the scaling symmetry of the reduced action may not correspond to the transformation corresponding to the on-shell path in the black hole  parameter space, we could manage to connect the conserved charge, $C$ associated with the scaling symmetry to  the physical quantities, the mass and/or the entropy. This is achieved  by computing the physical quantities by choosing the `on-shell' scaling transformation  and comparing them with the conserved charge $C$. 

In order to derive the Smarr relation, one needs to have the consistent description of the mass of black holes. Though it has been known that  boundary conditions of scalar fields may lead to non-integrable form of the infinitesimal mass, we have avoided such complicated situations  by restricting ourselves to hairy black holes with the definite on-shell scaling property.  This  boundary condition is implicitly taken care of  by choosing the definite `on-shell' scaling  transformation. 

Concretely, we have studied various hairy and non-hairy black holes in NMG and EMD gravity, whose asymptotic geometry is the one of the AdS space or the Lifshitz space. We find that the Smarr relation on  scalar hairy planar black holes is, generically, given by
\[
   M_{ADT}  = \frac{D-2}{D+z-2}T_H \CS_{BHW}\,,
\]
which is verified through various examples.
This form of the Smarr relation corresponds to the cases without the scalar chemical potential in the first law of black hole thermodynamics.  

It would be very interesting to realize black holes with more general boundary conditions of scalar fields and derive the Smarr relation in such cases. It would be also interesting to understand how to incorporate in our formalism the cosmological constant as a thermodynamic variable~\cite{Kastor:2009wy} in  the Smarr relation. One of the remaining interesting questions is the holographic interpretation of the scaling symmetry and its associated charge. In the view point of the AdS/CFT correspondence, the radial invariant charge $C$ means that the existence of RG flow invariant quantity in the field theory dual to the hairy black holes.

\vskip 1cm
\centerline{\large \bf Acknowledgments}
\vskip0.5cm
{SH was supported by the National Research Foundation of Korea(NRF) grant funded 
by the Korea government(MOE) with the grant number NRF-2013R1A1A2011548. JJ
was supported by the research grant ``ARISTEIA II", 3337 ``Aspects of three-dimensional
CFTs", by the Greek General Secretariat of Research and Technology.}

\vskip 1cm
\centerline{\large \bf Appendix}
\vskip0.5cm

\section*{A. Some useful formulae}
\renewcommand{\theequation}{A.\arabic{equation}}
  \setcounter{equation}{0}
In this appendix, we provide some detailed expressions of EOM and quasi-local ADT potentials, relevant in Sec. 3.
The scalar EOM for the given ansatz in NMG with the additional scalar field becomes 
\begin{align}
f \varphi''  + \bigg( A'f+\frac{f}{r} +f' \bigg)\varphi'
 + \frac{\alpha \varphi}{r} e^{-A}\bigg( 2(e^A f)' +r \Big( (e^A f)'+(e^A)'f \Big)' 
  \bigg) 
-\frac{\partial V(\varphi)}{\partial \varphi} =0 \,.
\end{align}
Two independent metric EOM are 
\begin{align}
\CE^{t}_{~t}- \CE^{r}_{~r}= &-\eta\frac{A'}{r}+\frac{\f'^2}{2} + \a \Big( \frac{A'\f^2}{2r} -e^A (e^{-A}\f\f')' \Big) \nn\\
  &+ \frac{\eta }{4r m^2}\bigg[ 2f \Big(A'^3 - 2A'^2 \Big(\frac{1}{r}-r A'' \Big) -A'(8A''+rA^{'''}) -(2rA''^2+3A^{'''}+rA^{''''})\Big) \nn\\
&\qquad\qquad +f'\Big(2rA'^3-16A''-7rA^{'''} -7A'^2 -3rA'A'' \Big)
- f'' \Big(11A' + 8rA'' - rA'^2\Big)  \nn\\
&\qquad\qquad - 2f^{'''} \Big( 1+rA' \Big) - rf^{''''}  \bigg]\,,
\end{align}
\begin{align}
\CE^{\theta}_{~\theta}& =  2\eta\L+\frac{1}{2}f\f'^2+ V(\f) +(\eta-\frac{1}{2}\a\f^2 ) e^{-A}\Big( (e^A f)' +(e^A)'f \Big)' 
-2\a e^{-A}\Big( e^A f \f\f' \Big)' \nn\\
 &~~~ + \frac{\eta}{8r^3m^2} \bigg[ r^2\Big( rA'^2 f'^2-2A'f'(8f'+3rf'')+r (-20f'^2A''+f''^2-4f'f^{'''}) \Big) \nn\\
&\qquad\qquad +4f^2 \Big( -4r^2A'^3+r^3A'^4+A'^2(5r-2r^3A'') -2A'(2+5r^2A''+3r^3A^{'''}) \nn\\
&\qquad\qquad\qquad~~ -r(-4A''+3r^2A''^2+2r A^{'''}+2r^2A^{''''}  ) \Big) \nn\\
&\qquad\qquad +4rf \Big( r^2A'^3f'-4rA'^2(3f'+rf'') -2A'( -4f'+7r^2A''f' +2rf''+2r^2f^{'''} ) \nn\\
&\qquad\qquad\qquad~~ -rf'(8A''+9rA^{'''})+r^2(7A''f''+f^{''''}) \Big)  \bigg]\,.
\end{align}
By applying the quasi-local ADT formalism, 
one can obtain quasi-local ADT potentials as
\begin{align} 
&\sqrt{-g} Q^{rt}_{ADT}(\xi_T ; \d g)|_{r\rightarrow\infty}   \nn \\
\!\!\!\!\!\!\! =&~ \frac{e^A}{2}  \bigg[ -\eta\d f+ \a(r\f^2)'\d f  \nn \\
&\qquad + \frac{\eta}{4 m^2} \Big( -\frac{8A'f}{r}+14 A'^2f -4rA'^3 f +3A'f'  +4A''f +8rA'A''f +3rA''f'-f''\nn\\
&\qquad \qquad \qquad  + 3rA'f'' +6rfA^{'''} +rf^{'''}\Big)\d f \nn\\
&\qquad +\frac{\eta}{m^2}\Big( 3A'f -\frac{1}{2}rA'^2 f+\frac{f'}{2} -\frac{3}{4}rA'f'+\frac{5}{2}rA''f \Big)\d f' \nn\\
&\qquad +\frac{\eta}{m^2}\Big( rA'f-\frac{1}{4}rf' \Big)\d f''+ \frac{\eta}{2m^2}rf \d f^{'''} \nn\\
&\qquad +\frac{\eta}{m^2}\Big( -\frac{f^2}{r} +5f^2A'-2rA'^2f^2+\frac{5}{2}ff'-\frac{1}{2}rA'ff' -\frac{3}{4}rf'^2+2rA''f^2 +\frac{3}{2}rff'' \Big)\d A' \nn\\
&\qquad + \frac{\eta}{m^2} \Big( f^2+rA' f^2+2rff' \Big) \d A'' + \frac{\eta}{m^2} rf^2\d A^{'''}\nn\\
& \qquad   -2\a r ( f'\f -2f\f'+2f\f A')\d\f +4\a rf\f\d\f'  \bigg]_{r\rightarrow\infty} \,,  \label{Qadt} \\
\nn\\
& \sqrt{-g} Q^{rt}_{ADT}(\xi_T ; \d \varphi)|_{r\rightarrow\infty}  = - \frac{1}{2}r e^A f \, \varphi' \, \delta\varphi  |_{r\rightarrow\infty}\,.
\end{align}
%
%

\newpage





\end{document}